# Dynamic disorder, phonon lifetimes, and the assignment of modes to the vibrational spectra of methylammonium lead halide perovskites

Aurélien M. A. Leguy[1], Alejandro R. Goñi[2,3*], Jarvist M. Frost[4**], Jonathan Skelton[4], Federico Brivio[4], Xabier Rodríguez-Martínez[2], Oliver J. Weber[4], Anuradha Pallipurath[5], M. Isabel Alonso[2], Mariano Campoy-Quiles[2], Mark T. Weller[4], Jenny Nelson[1,6], Aron Walsh[4] and Piers R.F. Barnes[1***]

1- Physics department, Imperial College London, UK, SW7 2AZ;
2- Institut de Ciència de Materials de Barcelona (ICMAB-CSIC), Campus UAB, 08193 Bellaterra, Spain;
3- ICREA, Passeig Lluís Companys 23, 08010 Barcelona, Spain ;
4- Chemistry department, University of Bath, UK, BA2 7AY;
5- School of chemistry, National University of Ireland Galway, Ireland;
6- SPECIFIC, College of Engineering, Swansea University, Baglan Bay Innovation and Knowledge Centre, UK, SA12 7AX;

\*       goni@icmab.es
\*\*     j.m.frost@bath.ac.uk
\*\*\*   piers.barnes@imperial.ac.uk

**Abstract**

We present Raman and terahertz absorbance spectra of methylammonium lead halide single crystals (MAPb$X_3$, $X$ = I, Br, Cl) at temperatures between 80 and 370 K. These results show good agreement with density-functional-theory phonon calculations.[1] Comparison of experimental spectra and calculated vibrational modes enables confident assignment of most of the vibrational features between 50 and 3500 cm$^{-1}$. Reorientation of the methylammonium cations, unlocked in their cavities at the orthorhombic-to-tetragonal phase transition, plays a key role in shaping the vibrational spectra of the different compounds. Calculations show that these dynamics effects split Raman peaks and create more structure than predicted from the independent harmonic modes. This explains the presence of extra peaks in the experimental spectra that have been a source of confusion in earlier studies. We discuss singular features, in particular the torsional vibration of the C-N axis, which is the only molecular mode that is strongly influenced by the size of the lattice. From analysis of the spectral linewidths, we find that MAPbI$_3$ shows exceptionally short phonon lifetimes, which can be linked to low lattice thermal conductivity. We show that optical rather than acoustic phonon scattering is likely to prevail at room temperature in these materials.

**Conceptual advances**

The hybrid perovskite family of compounds CH$_3$NH$_3$Pb$X_3$ (where $X$ = I, Br, Cl) can be used to make efficient solar cells and other semiconductor devices from low-cost ingredients deposited from solution. However, many fundamental properties of these materials have not been established, and, typically, devices made with them are not stable. We have compared the microscopic vibrations predicted in these crystals by *ab initio* calculations to peaks observed in Raman and terahertz spectra of CH$_3$NH$_3$Pb$X_3$ over a wide range of temperatures. Theory was consistent with observation, enabling us to comprehensively assign the complicated vibrational modes to spectral features, many of which are common to all three materials. For CH$_3$NH$_3$PbI$_3$, these features are no longer well resolved at temperatures above ~160 K, which can be explained because the organic moieties in the lattice cavities which are trapped in a given orientation in the low temperature phase is unlocked at temperatures above the phase transition. These insights into the types of active vibrations will help underpin our understanding both heat and electrical transport in these materials. Furthermore, the assignment of spectral peaks to particular vibrations which could be influenced by chemical changes will allow the stability of the materials exposed to different operating environments to be monitored using vibrational spectroscopy.

**Introduction**

Raman spectroscopy can characterize the chemical environments in materials *in situ*, as well as revealing the nature of the lattice vibrations (phonons). Solar cells based on methylammonium lead halide (MAPb$X_3$, $X$ = I, Br, Cl) have received considerable attention over the past few years, reaching power conversion efficiencies in excess of 20 % in 2015.[2] A precise understanding of the phonon dispersion in these materials is crucial for developing quantitative models of ionic transport,[3] and the recombination[4] and scattering[5] of (photo-generated) charges in devices. Reliable assignment of the optically accessible vibrational modes in MAPb$X_3$ ($X$ = I, Br, Cl) will also enable the development of high throughput spectroscopic methods to assess the perovskite film quality during the manufacturing process.

As yet, it has been difficult to fully interpret the Raman spectra of MAPbI$_3$. The material is extremely sensitive to focused laser illumination, which can cause irreversible degradation; Ledinsky *et al.*[6] were the first to demonstrate the necessity of sub-bandgap illumination to achieve sufficient signal while avoiding chemical changes to the material.

Subsequent Raman studies of MAPbI$_3$ using bias white LED light in addition to above-bandgap excitation indicated that reversible changes in spectra occurred on prolonged exposure of the material to illumination. Specifically, at room temperature, peaks in the Raman spectrum appeared between 50 and 200 cm$^{-1}$ after sustained illumination. These were attributed to a reversible change in the crystal structure.[7]

A first attempt at assigning the Raman spectra of MAPbI$_3$ to theoretical vibration modes at low wavenumbers was made by Quarti *et al.*;[8] however, the data acquired in this study are likely to contain contributions from degraded material induced by the optically-absorbed excitation source.[6] A tentative assignment of the Raman modes of MAPbCl$_3$ was proposed by Maaej and co-workers in Ref.[9], while studies of the Raman spectrum of the vibrations of the CH$_3$NH$_3^+$ cation can be found in Ref.[10] Together with various interpretations of the vibrational modes of related layered perovskite[11-14] and intercalation compounds,[15, 16] these studies provide solid basis for unambiguous assignment of the spectra of MAPb$X_3$. Recently we have published a preliminary Raman spectrum of MAPbI$_3$ in the orthorhombic phase, compared to first-principles lattice dynamics calculations on each of the three phases.[1] These calculations indicated that, at energies below 200 cm$^{-1}$, the spectrum can be attributed to a combination of the Pb/I cage modes and associated coupled motion of the CH$_3$NH$_3^+$ cations. In contrast, at wavenumbers greater than 200 cm$^{-1}$ only molecular vibrations of the cations contribute to the spectral features. This is as would be expected by the large mismatch in mass between the organic and inorganic components.

Infrared absorption spectroscopy complements Raman: Raman inactive vibrations are often found to be infrared active, and vice versa. Infrared absorption spectroscopy is also not subject to the same degradation issues as Raman; however, care must be taken to ensure artefacts are not introduced during sample preparation and mounting using halide salts (*e.g.* pelleting with KBr), which may undergo ionic exchange with the sample. Fourier-transform infrared (FTIR) spectra were recently reported for MAPbI$_3$ thin films deposited by dual evaporation[17] in the 6-3500 cm$^{-1}$ spectral range. The authors of Ref. [18] noted the presence of extra peaks in their spectra that could not be explained by any of the predicted modes, which they tentatively assigned to second order harmonic effects. Similar data was reported between 800-3600 cm$^{-1}$ for the three halide types at room temperature, noting few differences between them.[18] It is likely these features can be attributed primarily to vibrational modes of the CH$_3$NH$_3$ cation, as shown below and in Refs [17, 18]. Recent terahertz absorption spectra (overlapping the range of FTIR frequencies) have shown the existence of two vibrational bands around 1 and 2 THz (33 and 67 cm$^{-1}$, respectively), which were suggested to be coupled to the scattering (and thus mobility) of free carriers in the material.[5]

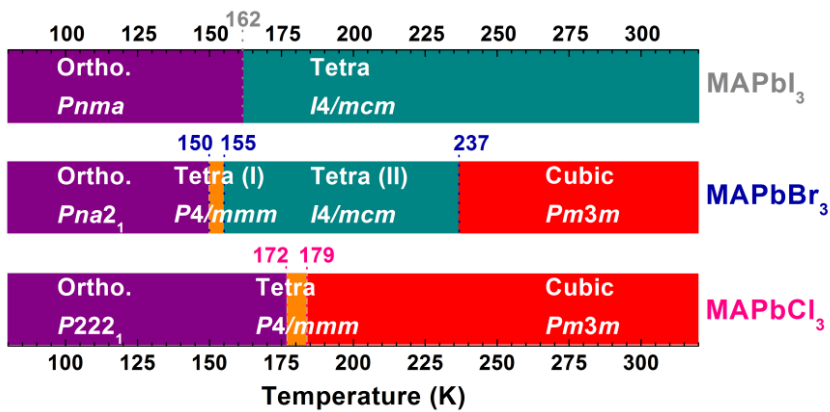

**Figure 1** Summary of the crystal systems and space groups adopted by MAPbI$_3$,[19-23] MAPbBr$_3$[20-23] and MAPbCl$_3$[9, 20-23] as function of temperature. The orthorhombic phases of each compound are represented in purple, the tetragonal phases of the space group *I*4/*mmm* and *I*4/*mcm* are in orange and dark cyan, respectively. Red is for the *Pm3m* cubic phase. The colour code shown in this figure is consistently kept throughout the paper.

The three compounds of the MAPb$X_3$ family ($X$ = I, Br, Cl) can exist in three different phases (four in the case of MAPbBr$_3$) as summarized in **Figure 1** for temperatures up to 325 K. Here, building on our preliminary work on MAPbI$_3$,[1] we assign the features of Raman and terahertz absorption spectra measured at a wide range of temperatures to their respective vibrations on the basis of first principles calculations of the phonon spectra of each of the compounds. This allows us to confirm our previous predictions of the Raman and FTIR modes expected in the three halide types. Our measurements also reveal that peaks in the vibrational spectra are strongly broadened (with the appearance of new broad peaks) as reorientation of the methylammonium cation becomes unlocked at the transition from the orthorhombic to tetragonal phase. We propose an explanation of the peculiar dependence of the C-N torsion mode on the cavity size that is a marker of the degree of steric hindrance experienced by the cations. Our calculations also show that additional peaks are expected from dynamic disorder. We then discuss the effects of phonon-phonon coupling on materials properties such as charge carrier mobility, the rate of hot carrier relaxation, and the implications for thermal conductivity.

**Methods**

*Sample preparation*

Single crystals of MAPb$X_3$ (MA = CH$_3$NH$_3$; $X$ = I, Br, Cl) were synthetized both from aqueous solution, based on the method described by Poglitsch and Weber,[20] and from inverse temperature crystallization in gamma-butyrolactone.[24] The detailed procedures are given in the **supplementary Note S1**.

*Raman spectroscopy*

Raman spectra were collected in backscattering geometry with a high resolution LabRam HR800 spectrometer using a grating with 600 lines per millimetre and equipped with a liquid-nitrogen cooled charge coupled device (CCD) detector. Raman measurements were carried out between 90 and 300 K using a gas flow cryostat with optical access that fits under the microscope of the Raman setup. Four optical sources are available for excitation: the 785 nm (infrared, IR) line of a diode-pumped solid-state laser, the 488 nm (blue) and 514.5 nm (green) lines from an $Ar^+$-ion gas laser, and the 632.5 nm (red) line from a He-Ne gas laser. The laser beam was focused onto the sample using a long working distance 20× microscope objective, yielding a spot size with a diameter of ~18 μm (area ~2.5 x $10^{-6}$ $cm^2$). Measurements of the cut-off energies of the different filters gives limits to the acquisition of spectra at low energies of ~20 $cm^{-1}$ for the IR laser, 50 $cm^{-1}$ for the red source, and 90 $cm^{-1}$ for the blue one. The measured power densities at the sample position for IR, red and blue irradiation were 310 $W/cm^2$, 70 $W/cm^2$ and 230 $W/cm^2$, respectively.

Heating by laser light directly absorbed by $MAPbI_3$ has been shown to lead to rapid degradation of the material, resulting in $PbI_2$ Raman signatures.[6] Since 785 nm light is only weakly absorbed, the heating effect of the laser was low enough to ensure the $MAPbI_3$ crystal structure was preserved. Samples were kept under vacuum inside the cryostat during the measurements. No degradation was observed for $MAPbBr_3$ and $MAPCl_3$ with any of the optical sources. Parasitic photoluminescence can also further limit the range of excitation energies, causing the Raman modes to disappear against a strong PL signal if they overlap. In practice, this was found only to be a significant issue for $MAPbBr_3$ at low energy. The IR laser was therefore used as excitation source for $MAPbBr_3$ below 1800 $cm^{-1}$, while the red source gives better results above this energy. $MAPbI_3$ and $MAPbCl_3$ are studied in the non-resonant regime across the whole spectral range (20-3500 $cm^{-1}$).

*Terahertz absorption spectroscopy*

To complement the Raman measurements on $MAPbI_3$, terahertz time-domain transmission spectra were recorded on this material between 0.2 to 5 THz, using an Advantest TAS7500TS system at a resolution of 7.6 GHz and 8192 integrated scans. A description of the measurement setup can be found in Ref. [25]. Samples were prepared by geometrically mixing crushed single crystals of MAPbI$_3$ (40 mg) with polytetrafluoroethylene (PTFE; 600 mg) powder and making pellets using a two-ton tablet press. A cuvette containing the pellet was attached to the cold finger of a continuous-flow cryostat and cooled at a rate of approximately 25 K min$^{-1}$ to 80 K. The system was placed under vacuum (2.8 x 10$^{-1}$ mbar). The temperature was controlled using a 50 W heater and data was acquired at variable intervals up to 370 K. The temperature was stepped by 10 to 20 K away from the phase transitions, while smaller 5 K measurement steps were performed close to 160 and 330 K.

*Computational methods*

The vibrational frequencies for each phase of MAPbI$_3$ and for the cubic phase of MAPbBr$_3$ and MAPbCl$_3$ were calculated within the harmonic phonon approximation using second order force constants obtained from density functional theory (DFT). We used the PBEsol density functional in the Vienna *ab initio* Simulation Package (VASP) code,[26-28] with rigorous convergence criteria to optimise the cell volume, shape and atomic positions, as described previously.[1] Spin-orbit coupling was not included as it does not influence the interatomic interactions at equilibrium; the conduction band formed by Pb 6p is unoccupied. The Phonopy package[29-31] was used to setup and post process supercell finite displacement calculations. The infrared (IR) activity was calculated using the mode eigenvectors and the Born effective-charge tensors. The Raman activity of each mode was obtained by following the eigenvector, and calculating the change in polarizability with an additional electronic structure calculation. Details are described in our earlier computational paper.[1]

Custom codes were written to provide eigenmode-resolved phonon partial densities of states; and to analyse the motion and energetic contribution of atoms at the gamma point.

Phonons are calculated around a local minimum on the potential energy surface of a material. The force constants, which define the change in force on a reference atom in response to the displacement of another, are used to construct a dynamical matrix, which is then diagonalised to give the eigenvalues (vibrational frequencies) and associated eigenvectors (normal modes of the motion). For a hybrid material, the potential energy surface has a complex structure, and there are multiple local potential energy minima. In hybrid halide perovskites, these minima are close in energy, such that thermal motion is sufficient to lead to continuous dynamic disorder. In order to sample the disorder in the phonon spectrum, we generated unit-cell structures with the methylammonium cation randomly orientated and displaced from its energy minimised location (see **supporting Note S2**). These structures were then energy minimised by up to 101 optimisation steps with a conjugate gradient algorithm. The zone-centre phonon spectrum was subsequently recalculated using density functional perturbation theory. Some of these structures exhibited imaginary phonon modes, which is expected for dynamic structure snapshots away from the ground state configuration. However, most presented a full set of positive frequency modes, indicating that they were in a local potential energy minimum, at least with respect to the local structure within the unit cell.

**Results and discussion**

*Characteristic features of the Raman spectra*

The measured spectra of the three halide types, *viz*. $MAPbI_3$, $MAPbBr_3$ and $MAPbCl_3$, are shown in **Figure 2**. The spectra can be understood as consisting of two parts: a low energy band of closely packed Raman peaks between the cut-off frequency of the optical filter at ~20 $cm^{-1}$ (the cut-off frequency for the THz IR measurements corresponds to 7 $cm^{-1}$) and ~200 $cm^{-1}$, and the higher energy Raman-active modes forming bands spread between ~200 and 3300 $cm^{-1}$.

As detailed in the **methods section**, particular care has to be taken when recording Raman spectra of $MAPbI_3$. The compound is extremely sensitive to focused laser light with above-bandgap frequency, as previously reported.[6] The use of above-bandgap lasers (488 nm, 514 nm or 633 nm) burns holes in the crystal, the result of which is a significant change in the collected spectra. This problem was avoided by using only sub-bandgap irradiation (785 nm) to excite $MAPbI_3$.

Parasitic photoluminescence can dominate the signal for particular excitation wavelengths, thus further limiting the range of usable laser wavelengths, as explained in the **methods section**. Moreover, different excitation wavelengths couple well with different vibrational modes. Short wavelength excitation allows the molecular modes at Raman shifts above ~ 200 $cm^{-1}$ to be clearly resolved, but is less suitable for resolving the cage modes at lower energies. Conversely, it is not possible to resolve the high wavenumber molecular modes with IR laser excitation (as shown for $MAPbBr_3$ in **Figure S1** in the **supplementary information**), yet this wavelength offers a clear picture of the inorganic sublattice (cage) modes. This is one of the reasons why it was not possible to obtain clearly resolved molecular peaks for MAPbI$_3$, since only IR irradiation could be used. The other reason for the lack of resolution of these molecular features is dynamic broadening, which is discussed in detail below. A better compromise can be found for MAPbBr$_3$ and MAPbCl$_3$, for which the cage modes could be investigated with IR light while the high-wavenumber molecular modes could be resolved with visible irradiation. An IR laser was used for MAPbBr$_3$ for the data collected between 200 and 1800 cm$^{-1}$ to avoid parasitic photoluminescence.

*Ab initio phonon calculations.*

First-principles lattice-dynamics calculations were performed to model the vibrational modes together with their corresponding spectroscopic intensities; see the **methods section** and Ref. [1]. The energies and relative intensities of the dominant modes are plotted in **Figure 2**. A clear separation in energy between the low frequency modes associated with the inorganic (PbI$_3^-$)$_n$ network and high-frequency modes of the organic CH$_3$NH$_3^+$ cations is evident. This separation is consistent with the two-part structure observed in the experimental spectra. The two-fold degeneracy of the *E* molecular modes is broken when the ions are caged in the cavity formed by the inorganic sub-lattice, forming separate bands with slightly red- or blue-shifted peaks. The torsional degree of freedom is formally of A$_2$ symmetry, which in the C$_{3v}$ point group of the methylammonium molecule can be neither Raman nor IR active. As such, its presence in the experimental spectra shows a coupling of this mode into the motion of the inorganic cage.

*Assignment of the brightest Raman peaks*

**Figure 2** shows satisfactory agreement between the experimental spectra and the calculations, although shifts in energy are apparent in some cases. Qualitatively, the same number of bands can be observed at comparable energies. For all three compounds in the cubic phase, a low-energy band (between 0 and 200 cm$^{-1}$) is comprised of 18 eigenmodes. The first three are acoustic modes involving the translation of the entire lattice. The 15 remaining modes are cage-dominated optical modes, including 6 which are the overall translation and rotation of the organic cation coupled into the lattice motions (these would not be present for the molecular modes in a vacuum) (see **Tables 1 and 2**). At higher energy (above ~200 cm$^{-1}$), 6 non-degenerate A modes and 6 doubly-degenerate E modes are predicted for the methylammonium cation $CH_3NH_3^+$ (see **Table 3**), which correspond to 18 predicted modes in the cubic phase once the symmetry-breaking effect of the surrounding lattice is taken into account (we number these modes 19 to 36). The tetragonal and orthorhombic configurations correspond to lower-symmetry arrangements, causing further splitting of most of the bands. This increases the number of predicted modes from 36 to 144. However, the energies of the split bands remain close together, such that once effects of disorder are accounted for (discussed below), they cannot be separately resolved experimentally. For this reason most of our discussion will refer to assignments of the modes of the cubic phases. **Figure 3** shows the phonon dispersion, and the density of phonon states for each of the three compounds at room temperature (see caption for details). Negative frequencies (or 'soft modes') are seen for the acoustic modes at the R and M points in the Brillouin zone, suggesting the possibility of anti-ferroelectric tilting behaviour in all three compounds. It is interesting to note the spread of energies for the groups of modes (sets of modes with the same colour hue but varying tones) that would be degenerate in an inorganic cubic perovskite. This spread is indicative of the disorder introduced in the material by the anisotropic nature of the MA ion. The gradient in the modes (d$E$/d$q$) becomes steeper with smaller halide ions, indicating an increase in the crystal 'stiffness' (bulk modulus).

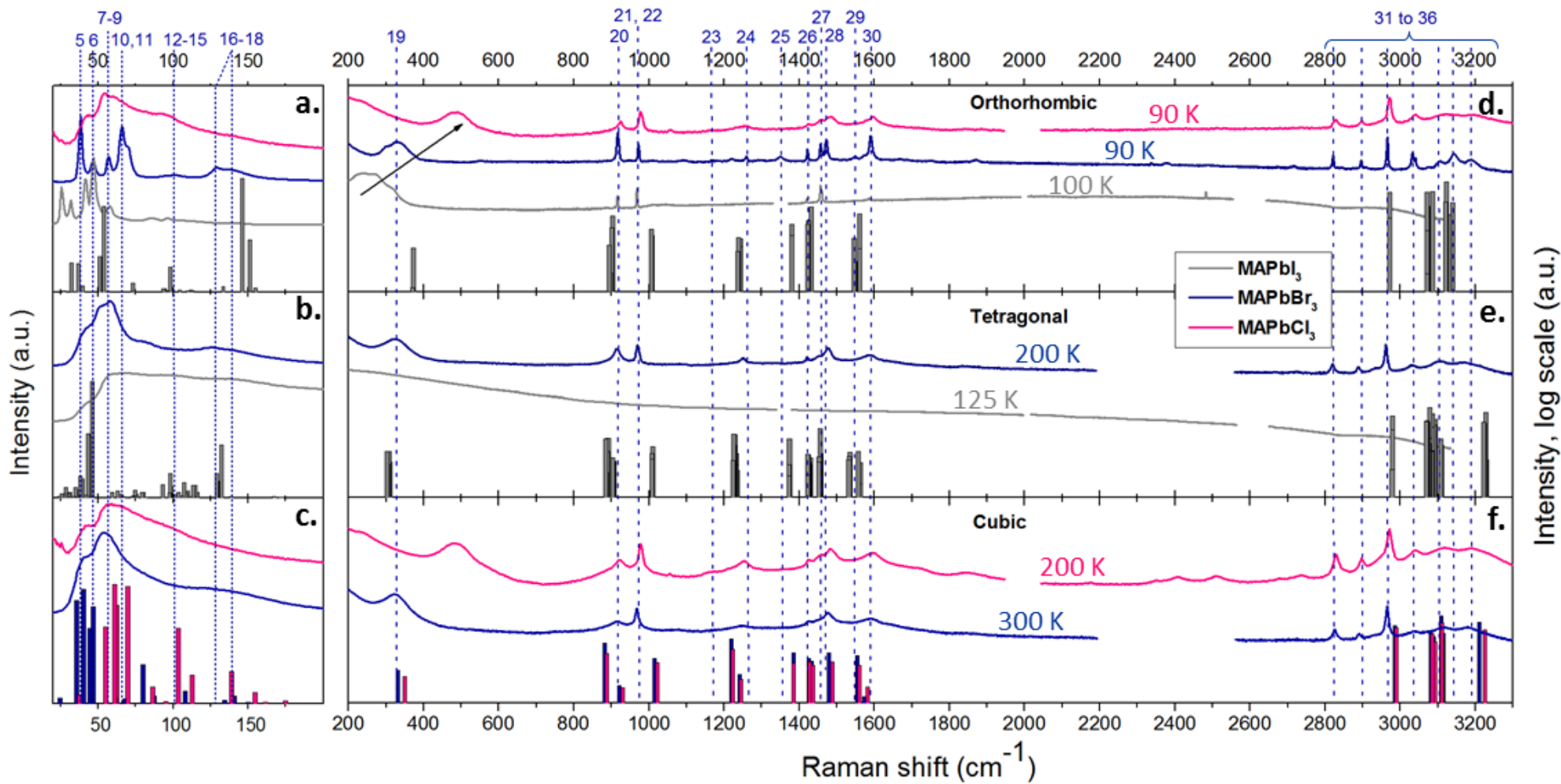

**Figure 2** Raman spectra of MAPbI$_3$ (solid grey line), MAPbBr$_3$ (solid blue line) and MAPbCl$_3$ (solid pink line) in their different phases: orthorhombic at low temperatures (top row, **a** and **d**), tetragonal at intermediate temperatures (middle row, **b** and **e**), and cubic at higher temperatures (bottom row, **c** and **f**). The temperatures are given above the spectra. The figure is split between the Raman shifts attributed to the cage modes (left hand side, **a** to **c**) and to the molecular modes (right hand side, logarithmic scale, **d** to **f**). The predicted eigenmodes of each phase are shown as bars in matching colours. The numbering of the eigenmode labels the main Raman peaks of MAPbBr$_3$ in the orthorhombic phase (defined in **Table 1**). Dotted lines show the positions of the labelled peaks and are intended as a guide to the eye. The arrow is used to emphasize the dramatic shift of mode number 19 for the different halides.

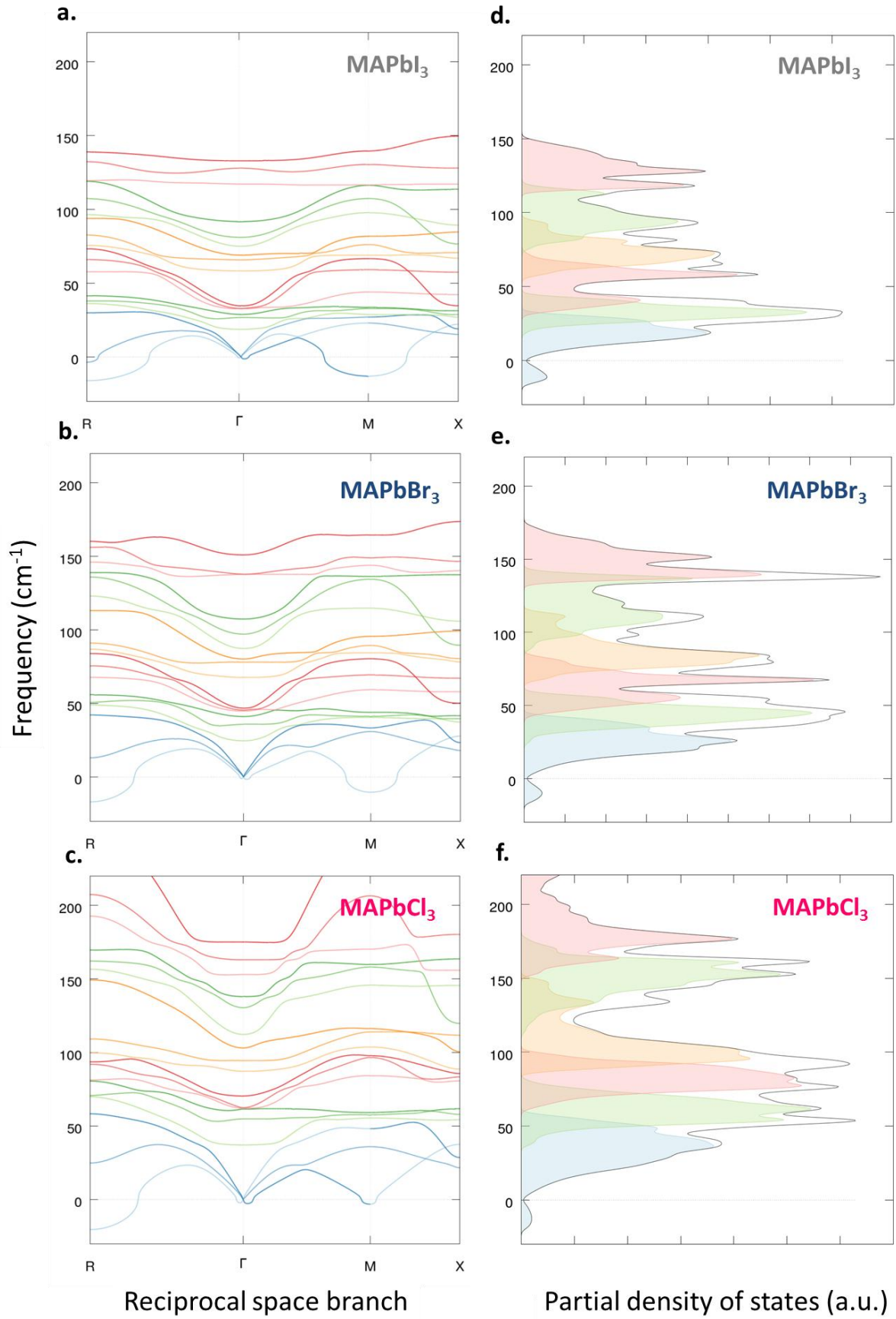

**Figure 3** Phonon dispersions of the 18 low frequency (inorganic cage) modes within the harmonic approximation using a pseudo cubic lattice for (**a**) MAPbI$_3$, (**b**) MAPbBr$_3$ and (**c**) MAPbCl$_3$ at room temperature. In all cases negative frequency, 'soft' modes are located around the Brillouin-zone boundary points $M[2\pi/a(1/2,1/2,0)]$ and $R[2\pi/a(1/2,1/2,1/2)]$. The band structure are plotted considering band crossings. The colour refers to the nature of the phonon eigenmode, which can change according to its proximity to other modes to avoid crossings (indicated by changes in shades). The three orthogonal acoustic modes (modes 1-3 in **table 2**) are plotted in blue shades. The remaining modes (4-18 in **table 2**) are optical, plotted in groups of three with a similar shade for each orthogonal mode (these would be degenerate if the MA ions were replaced by a spherical atom). Modes 4 – 6 are in green, 7 – 9 in red, 10 – 12 in orange, 13 – 15 in green and 16 – 18 in red. Density of phonon states decomposed by sets of three orthogonal phonon eigenmodes for MAPbI$_3$ (**d**), MAPbBr$_3$ (**e**) and MAPbCl$_3$ (**f**) integrated over the full Brillouin zone, but not considering band crossings.

*Assignment of the cage modes*

Our assignment of the experimental cage modes identified at 100 K for the three compounds (*i.e.* in their orthorhombic phases) is given in **Table 1** (see also **Figures S2-S4** in the **supplementary information**). The assignment is performed in the orthorhombic phase since the data is better resolved at low temperature, below the order-disorder phase transition, as will be discussed in detail below. Among the 144 modes expected in the orthorhombic phase, most have a small associated activity. Thus, only the brightest calculated peaks were considered for the assignment of the experimental features in the low symmetry phase. The assignment is interpreted by matching the predicted bright Raman modes in the orthorhombic phase to the nearest predicted band in the cubic phase (See **Figure S5**).

**Figure 4** shows the breakdown of the calculated eigenvectors in the cubic phase in terms of the relative energetic contribution of each element. For each of the three compounds, modes 1 to 18 show substantial involvement of the inorganic sublattice (cage) via both lead and halide-atom displacements. The effects of statistical order on these modes will be discussed following section. Modes 1, 2 and 3 are the acoustic modes, characterized by rigid translation of the lattice along the three Cartesian directions. These necessarily have zero frequency at the gamma point and, by their translational symmetry, no associated spectroscopic activity. Modes 4, 5 and 6 correspond to twists of the octahedra (see **Figure 5** for a graphical representation) that occur through the flexing of Pb-I-Pb bonds. These vibrational modes can be categorized as transverse-optic modes (TO), the eigenvector is symmetric with respect to inversion, and would be of $\Gamma_{25}$ symmetry in a fully inorganic cubic perovskite. The next three modes (7, 8 and 9) are the distortions of the octahedra by a change in angle of the I-Pb-I bonds, of $\Gamma_{15}$ symmetry. They exhibit a mixed nature of transverse- and longitudinal-optic phonons (TO and LO, respectively), the nature of which was inferred from the calculation of infrared activity. Modes 12 to 15 are truly coupled vibrations between cages and MA cations. They correspond to collisions of MA ions with the inorganic sub-lattice, correlated with changes in the Pb-I bond lengths. These modes are infrared active and LO-like.

The remaining cage modes stand out in **Figure 4**, since they show more balanced relative contributions from the inorganic and organic moieties. In particular, the 'nodding donkey' modes (10, 11, 16 and 18 in **Figure 5**) show moderate coupling between the cage and cation, with a smaller contribution from Pb atoms than for modes 12-15. They correspond to rotational vibration of the cation around the nitrogen (10 and 11) or the carbon atom (16 and 18). These motions are coupled with reorganizations of the surrounding octahedra, and would not be present in full inorganic perovskite. Similarly, mode 17 is the roll of the MA ions around their C-N axes, which is coupled to iodide displacements. It should be borne in mind that **Figure 4** is an energy breakdown of the modes, and so significant energetic contributions from the Pb and I atoms correspond to very small-amplitude displacements from equilibrium because of the high atomic mass relative to those of N, C and H.

For the iodide perovskite, complementary terahertz absorption-spectra measurements were carried out to investigate vibrational modes below the cut-off frequency of the Raman setup. **Figure 6** shows the spectra measured for MAPbI$_3$ at temperatures ranging from 80 to 370 K. To help identify the absorption features against the broad background, the second derivative of the terahertz absorbance is given in the **supplementary Figures S6-S8**, and compared with expected infrared-active vibrational modes obtained from the calculations. The experimental data are in good agreement with the predicted infrared activity and clearly shows two bands, centred around 1 and 2 THz, which is in agreement with previous observations.[5] The 1 THz band can be attributed to the octahedral twists and distortions, while the 2 THz band occurs where the nodding donkey around N and the "lurching" modes are observed. A complete assignment of the distinguishable features is given in **Table 2**.

Our observation of low energy modes is consistent with these materials behaving as soft semiconductors. This would imply that the material is potentially susceptible to ferro-elastic phase changes[32] in the presence of electrostatic forces induced by accumulations of separated charges in the material, which might explains recent observations of crystal photo-/electro-striction.[33] The strong variation in the width of the Raman modes observed with temperature (discussed further in the section on temperature evolution) implies that similar changes in Raman activity could be expected with variation in crystal structure induced by other mechanisms (such as those induced by changes in the electric field in the material resulting from an accumulation of photogenerated or ionic charges). A process such as this might be consistent with the Raman properties varying with illumination time observed by Gottesman et al.[34]

We note that formation and accumulation of iodide vacancies in the crystal lattice $MAPbI_3$ at room temperature might add structure to its Raman spectrum. Changes in the mass of the average oscillators involving iodine is expected to result in a blue-shift of the associated peaks. Modes become localised by the increased defect density, which disrupts the selection rules that prevent the observation of modes in perfect lattices, and therefore might reveal previously suppressed modes.

**Table 1**   Assignment of the Raman peaks of $MAPbI_3$, $MAPbBr_3$ and $MAPbCl_3$ associated with the cage vibrations in the orthorhombic phase (excitation wavelength 785 nm, samples at 100 K, "b" = broad, "sh" = shoulder, "LO" = longitudinal optic, and "TO" = transverse optic). The peak positions are compared to the brightest calculated Raman features of $MAPbI_3$ in the orthorhombic phase (see **Table S1** in the **supplementary information**). A graphical illustration of the assignment can be found in the **supplementary information**, **Figures S2-S4**. The peaks are tentatively related to those observed in the cubic phase (see also **Figure S5**). A schematic representation of the vibrational mode eigenvectors can be found in **Figure 5.**

| $f_{theo, MAPbI3}$ orthorhombic (cm$^{-1}$) | $f_{MAPbI3}$ exp 100 K (cm$^{-1}$) | $f_{MAPbBr3}$ exp 100 K (cm$^{-1}$) | $f_{MAPbCl3}$ exp 100 K (cm$^{-1}$) | Description | Corresponding mode number in the cubic phase (tentative) |
|---|---|---|---|---|---|
| **0** | - | - | - | Acoustic mode | 1 |
| **0** | - | - | - | Acoustic mode | 2 |
| **0** | - | - | - | Acoustic mode | 3 |
| **15** | - | - | 42 | Octahedra twist (TO) | 4 |
| **27** | 26 | 39 | 54 | Octahedra twist (TO) | 5 |
| **32** | 32 | 47 | 61 | Octahedra twist (TO) | 6 |
| **37** | 42 | 58 | 75 | Octahedra distortion (LO/TO) | 7, 8 |
| **39** | 47 | 58 | 75 | Octahedra distortion (LO/TO) | 9 |
| **51** | 58 | 66 | 94 | Nodding donkey around N | 10 |
| **54** | 58 | 66 | 94 | Nodding donkey around N | 11 |
| **73** | sh | 71 | 94 | Lurching MA (LO-like) | 12 |
| **93** | 86 | 99 | 124 b | Lurching MA (LO-like) | 13 |
| **94** | 86 | 99 | 124 b | Lurching MA (LO-like) | 13 |
| **98** | 97 b | 129 | 168 b | Lurching MA (LO-like) | 14 |
| **104** | 97 b | 129 | 168 b | Lurching MA (LO-like) | 14 |
| **112** | 97 b | 138 | 168 b | Lurching MA (LO-like) | 14 |
| **133** | sh | - | 184 | Lurching MA (LO-like) | 15 |
| **146** | 143 b | 148 | 238 b | Nodding donkey around C | 16 |
| **151** | 143 b | 148 | 238 b | Attempt roll around C-N | 17 |
| **155** | 143 b | 148 | 238 b | Nodding donkey around C | 18 |

**Table 2**  Assignment of the THz peaks of MAPbI$_3$ associated with the cage vibrations in the orthorhombic (80 K), tetragonal (220 K) and cubic phases (350 K) (see also **Figures S6-S8** in the **supplementary information**). In the table, "exp" = experimental, "w" = weak, "b" = broad, "sh" = shoulder, "s" = strong, "LO" = longitudinal optic, and "TO" = transverse optic. Features which are partially distinguishable, but not reliably so, are marked with a question mark. A schematic representation of the vibrational mode eigenvectors can be found in **Figure 5**.

| mode # | $f_{theo, MAPbI3}$ cubic (cm$^{-1}$) | $f_{MAPbI3}$ exp. 80 K (cm$^{-1}$) | $f_{MAPbI3}$ exp. 220 K (cm$^{-1}$) | $f_{MAPbI3}$ exp. 350 K (cm$^{-1}$) | Description |
|---|---|---|---|---|---|
| 1 | 0 | - | - | - | Acoustic mode |
| 2 | 0 | - | - | - | Acoustic mode |
| 3 | 0 | - | - | - | Acoustic mode |
| 4 | 19 | 21 sh | 22 sh | 19 sh, w | Octahedra twist (TO) |
| 5 | 27 | 27 sh | 26 sh, w | 28 sh, w | Octahedra twist (TO) |
| 6 | 29 | 33 | 29 sh, w | 31 sh, w | Octahedra twist (TO) |
| 7 | 33 | 35 s | 32 sh | 35 s | Octahedra distortion (LO/TO) |
| 8 | 33 | 35 s | 32 sh | 35 s | Octahedra distortion (LO/TO) |
| 9 | 35 | 38 s | 35 s | 35 s | Octahedra distortion (LO/TO) |
| 10 | 58 | 57 sh, b | 51 sh, b | 56 sh, b | Nodding donkey around N |
| 11 | 66 | 63? | 64 sh | 56 sh, b | Nodding donkey around N |
| 12 | 69 | 69 s | 72 s | 73 s | Lurching MA (LO-like) |
| 13 | 75 | 71 s | 76 s | 78 s | Lurching MA (LO-like) |
| 14 | 81 | 80 sh, w | 81 sh, b | 81 s | Lurching MA (LO-like) |
|  |  | 82 sh, w | 83 sh, b |  |  |
| 15 | 92 | 93 sh | 93 sh, w | 95? | Lurching MA (LO-like) |
| 16 | 117 | 106 b | 106? sh, b | 102? | Nodding donkey around C |
| 17 | 128 | - | - | - | Attempt roll around C-N |
| 18 | 133 | 147 sh | 141? | 138? | Nodding donkey around C |

*Assignment of the isolated mode*

While the majority of the molecular modes show no significant change in wavenumber between the different halides, the feature labelled eigenmode 19 in the spectra in **Figure 2** is atypical in that it exhibits a blue shift as the average lattice constant shrinks from ~6.3 to ~5.9 to ~5.7 Å for $X$ = I, Br and Cl, respectively[19, 20, 35] (corresponding to lattice volumes of 254, 206 and 183 Å$^3$).[20, 35] We attribute this 'isolated' band around 240 cm$^{-1}$ for MAPbI$_3$, 310 cm$^{-1}$ for MAPbBr$_3$ and 480 cm$^{-1}$ for MAPbCl$_3$ (see **Table 2**) to the torsional vibration of CH$_3$NH$_3^+$, which is consistent with previous reports.[9, 17, 36, 37] It is striking that this mode shows such high sensitivity to the halide atom in the compound, orders of magnitude more acute than in the other molecular modes. It is important to note that this mode is spectroscopically inactive outside the lattice – to be observed at all, the motion has to be coupled with the cage. **Figure 4** reveals that hydrogen atoms store more than 95% of the energy in this C-N torsion mode. The sensitivity of the MA torsional movements to steric hindrance by the surrounding lattice cages suggests that the spectral peak corresponding to this vibration could be used to probe variation of lattice structure in spatially-resolved studies, or in mixed-halide systems.

Similar behaviour is observed with mode 17, which corresponds to the rotation of the whole MA cation around the C-N axis. This mode is strongly populated in MD simulations[1] and corresponds to a large-amplitude movement upon which steric hindrance has a strong influence.

The interpretation of the 'isolated band' could be complicated by the possibility that the MA ions may have a disordered arrangement within the lattice. We have recently used quasi-elastic neutron scattering measurements to show that the MA ions jump between different orientations within the lattice; simulations of this process suggested increasingly disordered orientations with increasing temperature.[38]

The measured 'isolated' feature assigned to C-N torsional vibrations is broader and shows more structure than expected, but no other phonon mode is anticipated in the region. To further investigate this interesting feature, we performed additional calculations to take statistical disorder into account. A hundred snapshots were considered in which the organic cation was randomly oriented and displaced, and only allowed partial relaxation before calculating the vibrational features of the crystal. The effect of disorder on mode 19 in $MAPbI_3$ is shown in **Figure 7** (see **Figure S9-S11** for the complete set of calculated spectra of the three halide systems with dynamic disorder taken into account). The feature is changed in two major ways: (i) the resulting spectral feature, as a convolution of the 100 individual snapshots, is noticeably broadened compared to the ordered system. Full widths at half maxima (FWHMs) exceeding 150 cm$^{-1}$ are expected, which match well with the measured linewidths for this mode; (ii) a second peak is created at lower Raman shift energies (a shoulder in the case of $MAPbI_3$), illustrating that additional spectral features can result from disorder only. We believe that these extra peaks may have been misinterpreted in former studies as being harmonic overtones or combination bands, which would be much less intense and thus barely detectable. Statistical orientational disorder is likely to be intermixed with dynamic disorder. The effect of dynamic disorder on the peak widths is detailed in the section below discussing the temperature dependence of the Raman spectra.

*Assignment of the molecular modes*

Assigning the predicted molecular modes to the nearest Raman feature seems justified (see **Figures S12-S14** in the **supplementary information** for annotated spectra) and is consistent with previous reports on MAPbCl$_3$.[9, 10] The internal modes of CH$_3$NH$_3^+$ show little change upon halide substitution or across the phase changes. In fact, the peak positions are close to what is obtained for bare methylammonium ions.[10] This observation corroborates the FTIR investigation of Glaser *et al.*[18] Our tentative assignment is given in **Table 3**. Since little temperature evolution can be seen, the corresponding theoretical peaks are given for the cubic phase. **Figure 5** shows the eigenvectors associated with each vibrational mode. **Figure 4** shows that modes 20 to 36 are similar for the three halide types, and are composed almost entirely (i.e. >99 %) of displacements of the atoms in the cations. A collection of weak unidentified features can be observed between 1000 and 1200 cm$^{-1}$ as well as extra peaks above 2500 cm$^{-1}$ (see **Figure 2**). These features might be a consequence of stochastic disorder in the material. For the features above 2500 cm$^{-1}$, hydrogen atoms are likely to be significantly involved in the vibrations, and so these features might therefore correspond to different hydrogen-bonding configurations. We note however that even if the samples were kept in deep vacuum, we cannot exclude the possibility that the unidentified features are parasitic peaks from trapped solvent[38, 39] or due to water ingress into the material.[40, 41]

**Table 3**    Assignment of the molecular Raman peaks of MAPbI$_3$, MAPbBr$_3$ and MAPbCl$_3$ in the orthorhombic phase (see **Figures S12-S14** in the **supplementary information**). The excitation wavelength was 785 nm for MAPbI$_3$, 785 nm for MAPbBr$_3$ between 200-1700 cm$^{-1}$ and 633 nm above, and 488 nm for MAPbCl$_3$. The samples were measured at 100 K. In the table, "w" = weak, "b" = broad, "sh" = shoulder, "s" = strong, "LO" = longitudinal optic, "TO" = transverse optic and features marked "?" are those which are difficult to reliably detect. Since little temperature evolution is expected for the molecular Raman peaks, the comparison given with the theoretical modes of MAPbI$_3$ is to the cubic phase (see **Table S1** in the **supplementary information** for an extensive list of the 72 undegenerate molecular

modes of MAPbI$_3$ in the orthorhombic phase). See **Figure 5** for a graphical representation of the eigenvectors of the 18 molecular modes.

| mode # | Sym. group | $f_{theo, MAPbI3}$ cubic (cm$^{-1}$) | $f_{MAPbI3}$ (cm$^{-1}$) | $f_{MAPbBr3}$ (cm$^{-1}$) | $f_{MAPbCl3}$ (cm$^{-1}$) | Description |
|---|---|---|---|---|---|---|
| 19 | A$_2$ | 315 | 199 / 223 / 243 / 272 / 312 | 297 / 326 | 483 | C-N torsion |
| 20 | E | 876 | 889 | 915 | 925 | ? |
| 21 | E | 909 | 916 | 970 | 978 | C-N asymmetric bend |
| 22 | A$_1$ | 1007 | 968 | 994 | 1002 | C-N stretch |
|  |  |  |  | 1017 | 1058 |  |
|  |  |  |  | 1059 |  |  |
| 23 | E | 1215 | 1008? | 1115 | 1164 | C-N symmetric bend |
|  |  |  |  | 1143 |  |  |
|  |  |  |  | 1178 |  |  |
| 24 | E | 1234 | 1043? | 1236 | 1255 | C-N symmetric bend |
| 25 | A$_1$ | 1378 | - | 1352 | - | CH$_3$ asymmetric breathing |
| 26 | E | 1418 | 1420 | 1421 |  | CH$_3$ symmetric breathing |
| 27 | E | 1425 | 1420 | 1458 | 1425 | CH$_3$ symmetric breathing |
| 28 | A$_1$ | 1464 | 1457 | 1471 | 1457 | NH$_3$ asymmetric breathing |
| 29 | E | 1542 | sh | 1573 | 1485 | NH$_3$ symmetric breathing |
| 30 | E | 1558 | 1586 | 1590 | 1596 | NH$_3$ symmetric breathing |
| 31 | A$_1$ | 2971 | 2949b | 2821 | 2830 | C-H asymmetric stretch |
| 32 | E | 3067 | 2949b | 2896 | 2899 | C-H symmetric stretch |
| 33 | E | 3074 | 2949b | 2965 | 2972 | C-H symmetric stretch |
| 34 | A$_1$ | 3087 | 2949b | 3033 | 3040 | N-H asymmetric stretch |
| 35 | E | 3096 | 2949b | 3106 | 3117 | N-H symmetric stretch |
| 36 | E | 3199 | 2949b | 3144 | 3190 | N-H symmetric stretch |
|  |  |  |  | 3179 |  |  |

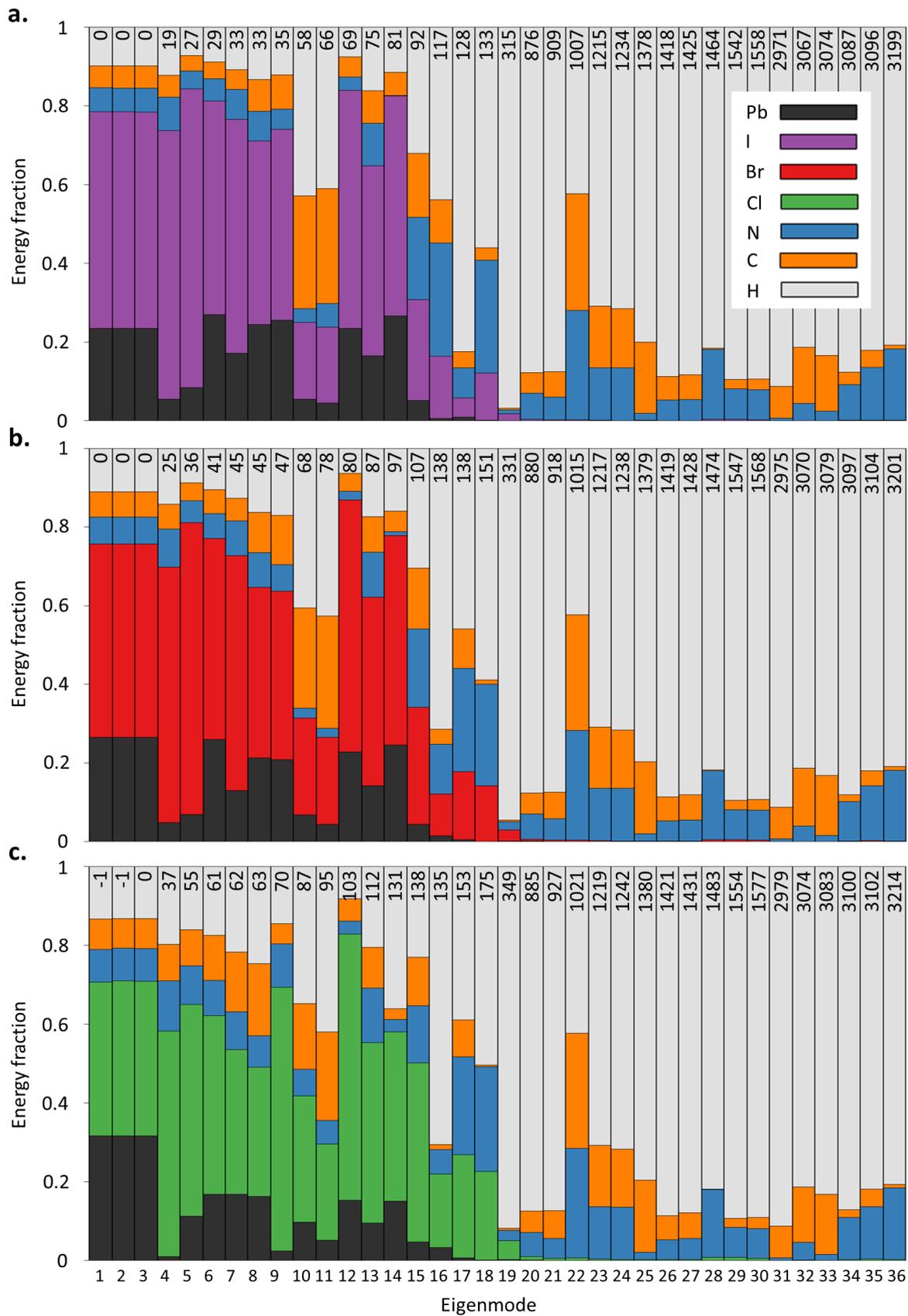

**Figure 4** Decomposition of the 36 gamma-point eigenmodes of cubic MAPbI$_3$ (**a**), MAPbBr$_3$ (**b**) and MAPbCl$_3$ (**c**) into relative energetic contribution of each atom. The contribution of Pb is shown in dark grey, C in orange, N in blue, H in white, I in purple, Br in red and Cl in green. The number of each mode is given at the bottom of the plot, and the corresponding frequency (in cm$^{-1}$) at the top.

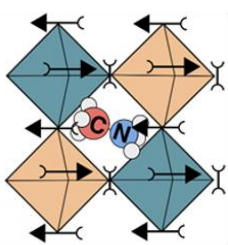 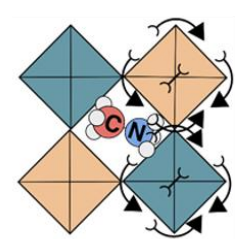 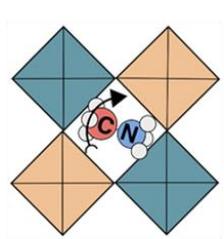 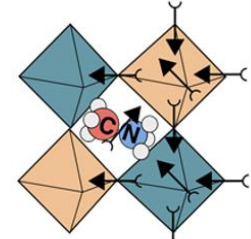 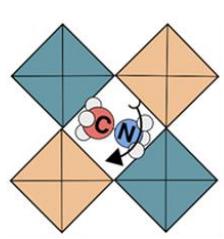 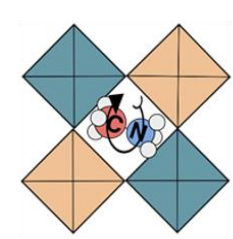

Modes: 4, 5, 6  TO     Modes: 7, 8, 9  LO/TO     Modes: 10, 11     Modes: 12-15  ~LO     Modes: 16, 18     Mode: 17

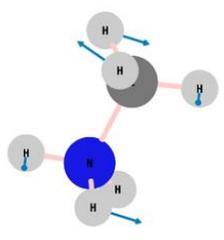 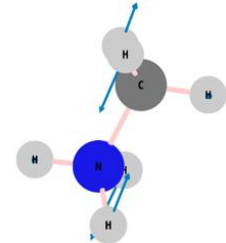 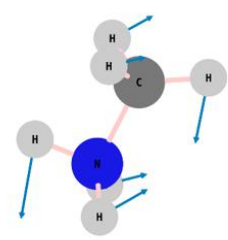 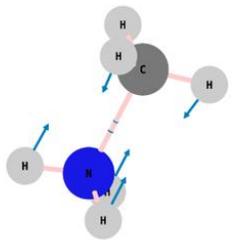 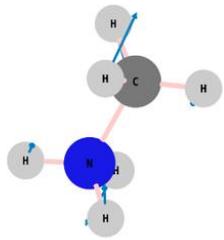 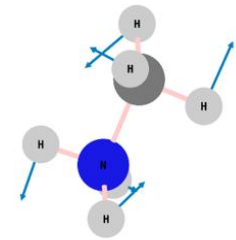

Mode: 19   Symm: A     Mode: 20   Symm: E     Mode: 21   Symm: E     Mode: 22   Symm: A     Mode: 23   Symm: E     Mode: 24   Symm: E

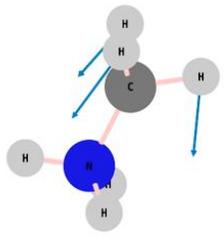 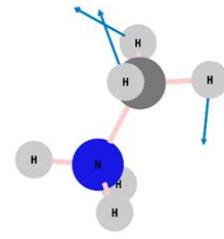 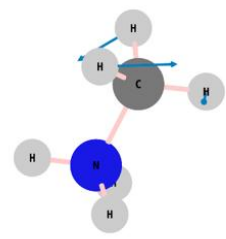 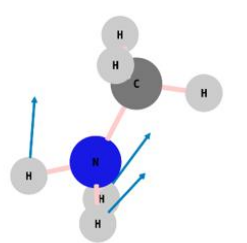 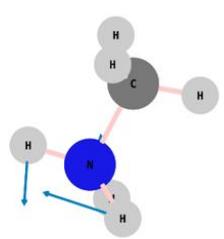 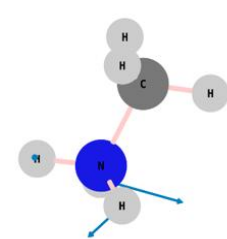

Mode: 25   Symm: A     Mode: 26   Symm: E     Mode: 27   Symm: E     Mode: 28   Symm: A     Mode: 29   Symm: E     Mode: 30   Symm: E

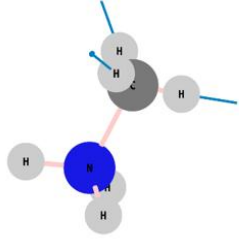 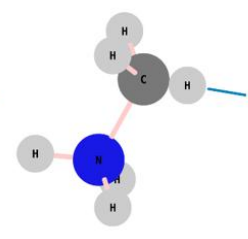 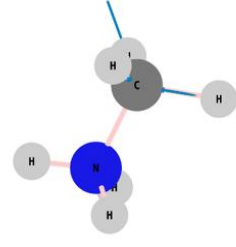 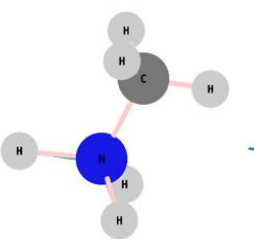 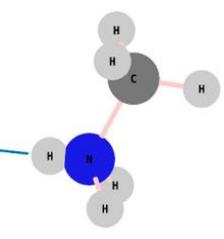 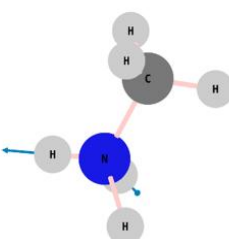

Mode: 31   Symm: A     Mode: 32   Symm: E     Mode: 33   Symm: E     Mode: 34   Symm: A     Mode: 35   Symm: E     Mode: 36   Symm: E

**Figure 5** Schematic representations of the vibrational modes geometries of the 36 zone-centre (Γ-point) phonon modes of the methylammonium lead halides in the cubic phase. The first row shows the 6 cage-vibration types (the acoustic modes are omitted, since they correspond to translations of the whole lattice and are therefore spectroscopically inactive). The last three rows show the molecular modes (numbered 19 to 36).

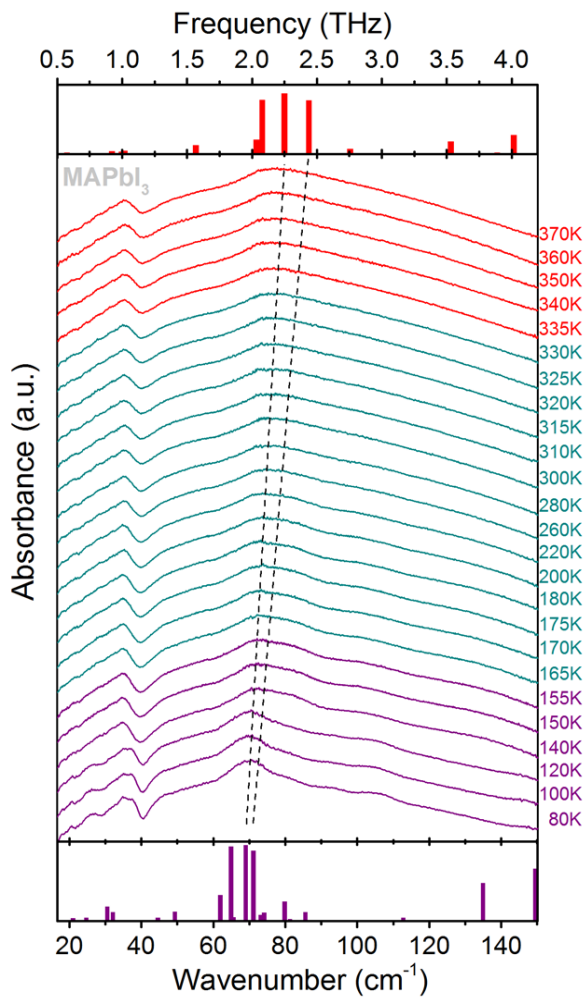

**Figure 6** Terahertz absorption spectra of MAPbI$_3$ measured at temperatures between 80 and 370 K. Spectra were recorded in the 0.5 to 4.5 THz energy range with 10 to 20 K steps away from the phase transitions, and 5 K steps around 160 and 330 K. The top and bottom bar charts give the predicted infrared activity of the low-frequency modes in the cubic and orthorhombic phases, respectively. The dotted lines join infrared-

active modes in the orthorhombic and cubic phases, and are intended as guides for the eye in order to visualise the shift and broadening of the band around 2 THz.

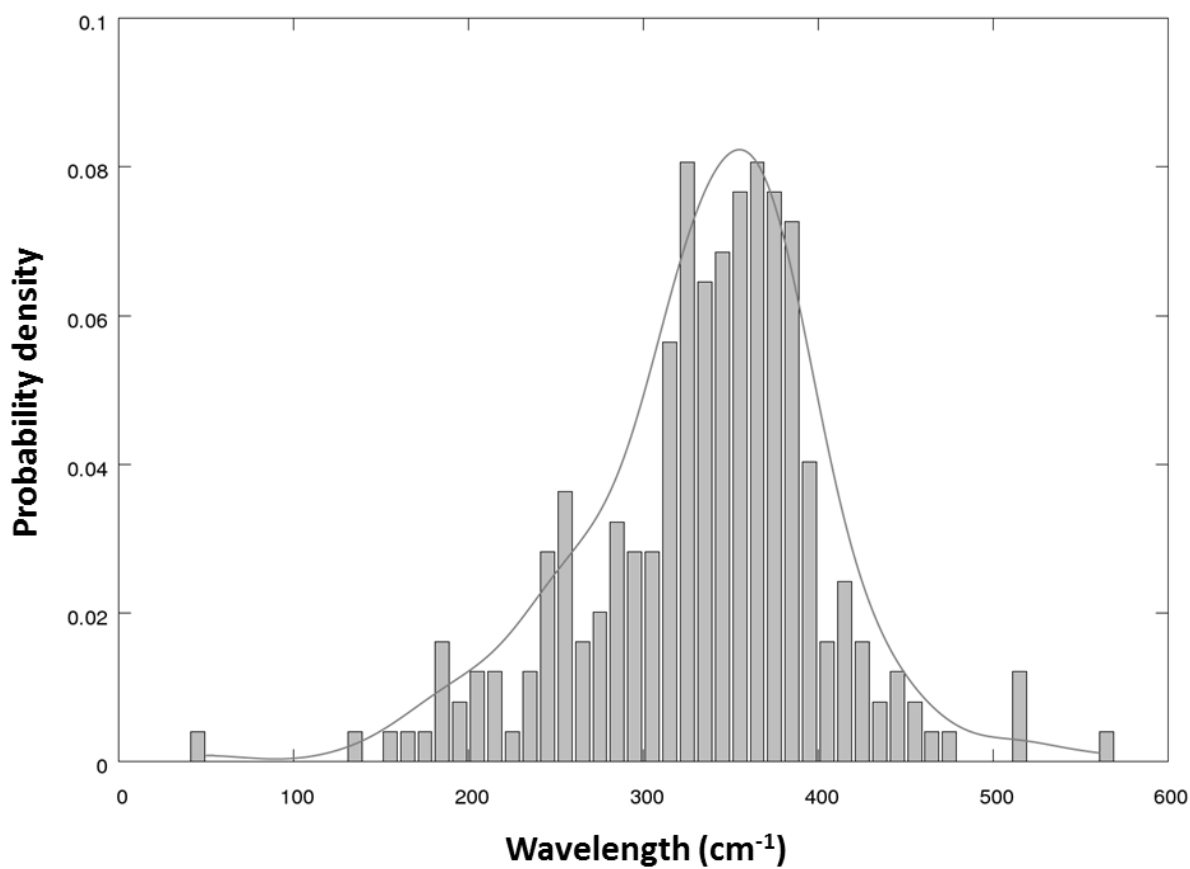

**Figure 7**   Effect of disorder on the frequency of the MA torsional mode (19 in **Figure 3** and **Table 3**) of MAPbI$_3$. The histograms show the distribution of frequencies of mode 19, extracted from 248 individual calculations in which the MA cation was randomly orientated and partially relaxed (see supplementary information). The solid line is the kernel density estimation of these sampled data.

*Temperature evolution of the modes*

In the case of MAPbI$_3$, **Figures 8a to 8e** show that most of Raman peaks weaken in intensity and broaden at the orthorhombic-to-tetragonal phase change (around 160 K, see **Figure S15 and S16** in **the supplementary information**). **Figure S17a** in the **supplementary**

**information** further illustrates this phenomenon by revealing a clear and sharp step-like increase of the peak linewidths when the temperature is increased across the low-temperature phase transition. The broadening is so strong in the case of $MAPbI_3$ that some Raman features almost disappear above ~150 K. For example, the sharp molecular doublet with peaks at 925 and 980 cm$^{-1}$ in **Figure 8d**, which we assigned to the C-N bending and stretching, respectively, disappears altogether across the phase change. We attribute this extreme broadening to the change of dynamics which occurs when the full reorientation of the methylammonium ions inside the cavity is unlocked.

The rapid realignment of the C-N axis of $CH_3NH_3^+$ was predicted in MD studies[42, 43] and observed by nuclear magnetic resonance (NMR),[44, 45] adiabatic calorimetric measurements,[45] quasi-elastic neutron scattering (QENS)[38, 46] and optical 2D vibrational spectroscopy.[47] However, there is no Raman peak directly ascribed to this motion, since it is correlated with cage vibrations. In other words, since the MA cations are very light compared to the surrounding lattice, we expect that they can be 'pushed around' by the cage vibrations. This creates a continuum of different environments for the cations, thus causing a strong broadening of most of the vibrational features. The unlocking of what is effectively a new dynamical degree of freedom creates a statistical distribution of vibrational modes corresponding to the resultant different local crystal configurations.

Further insight is obtained from the functional form that yields the best description of the spectral features. All the Raman peaks of the three compounds are best described by sharp Lorentzian functions in the low temperature orthorhombic phase. An increase of temperature in this phase leads to an increase in the peak width although the peak retains its Lorentzian shape (known as homogenous broadening). The homogeneous broadening occurs because the activation of constrained MA reorientations produces dynamic disorder which, on average, is distributed evenly through the lattice, this results in an increased spread in scattering energies. However, the nature of the temperature-induced broadening changes abruptly in the tetragonal phase. Some of the cage modes of $MAPbI_3$ become better fitted by Gaussians functions in this phase (see **Table S2** in the **supplementary information**). This is known as inhomogeneous broadening and is likely to be due to the appearance of a distribution of bond lengths resulting from spatial disorder in the orientation and/or position of the MA molecules within the perovskite cages. Inhomogeneous broadening is expected to be temperature-independent if the structure and dynamics of the material is conserved.

Inhomogeneous broadening of Raman peaks is a recurrent feature in disordered materials. It has been reported in manganese perovskites such as $LaMnO_3$ near the displacive orthorhombic-to-cubic phase change, and has been attributed to an increase in lattice disorder.[48] A comparable effect can occur with organic (*e.g. trans*-stilbene[49]) or inorganic (*e.g.* $SF_6$ and $WF_6$, Ref. [50]) molecules, and is often interpreted as statistical broadening due to environmental fluctuation. Temperature-independent inhomogeneous broadening has been reported for III-V semiconductors (*e.g.* GaAs, InAs, GaSb and InP, Ref. [51]), and is considered to be a surface effect caused by the microscopic disorder generated during the polishing process.

**Figure 8a**, shows the temperature evolution of the FWHM in $MAPbI_3$. The Raman features associated with cage vibrations merge across the phase change around 160 K, and the corresponding abrupt broadening of the peaks can be seen in **Figure S17a** in the **supplementary information**. Similarly, the terahertz absorption spectra given in **Figure 6** show significant broadening over the 80 to 370 K temperature range. In particular, the structure in the broad feature around 2 THz gradually disappears, presumably as the individual peaks broaden. The typical increase of FWHM of a cage mode in $MAPbI_3$ at the orthorhombic-tetragonal transition is from about $W_L$ ~10 to $W_G$ ~50 cm$^{-1}$ yielding the ratio $W_L/W_G$ ~5 ($W_L$ and $W_G$ are the FWHMs of Lorentzian and Gaussian peaks, respectively). Furthermore, **Figure S17a** in the **supplementary information** shows that $W_G$ in the tetragonal phase is temperature independent. At room temperature, broadening of the Raman peaks is dominated by inhomogeneous effects ($W_G \gg W_L$) that we attribute to a wide spread of bond-length distribution correlated with the variety of allowed MA orientations.

A broadening of the cage peaks is also apparent across $MAPbBr_3$ the Tetragonal I to Tetragonal II phase change, although this is less extreme than in the case of $MAPbI_3$. In particular, the individual modes can still be traced across the phase change, as can be seen in **Figures 8f to 8j**. For $MAPbBr_3$, the typical broadening of a cage mode is $W_G/W_L$ ~2. Despite the fact that the broadening in $MAPbBr_3$ is stronger for the cage modes, the molecular peaks exhibit a non-negligible broadening. Once in the tetragonal phase, there is still an increase in linewidth with increasing temperature, but this is by far less pronounced than at the phase transition. Similar abrupt broadening as in $MAPbI_3$ can be seen from **Figure S17b** in the **supplementary information**, coinciding with a change in the nature of the first five cage modes from Lorentzian to Gaussian (see **Table S2-S4** in the **supplementary information**).

Evidence of the reorientation of the cation in MAPbBr$_3$ being unlocked between 150 and 155 K can be found in the dielectric and calorimetric measurements reported by Onoda-Yamamuro.[21, 22] More recently, Swainson et al.[52] studied the relaxation dynamics of MAPbBr$_3$ by QENS, and confirmed the melting of orientation order at this temperature.

In the case of MAPbBr$_3$, $W_G$ and $W_L$ are similar in the disordered phase. Due to the structural similarities between MAPbI$_3$ and MAPbBr$_3$, we believe that the homogeneous component of the broadening $W_L$ in MAPbBr$_3$ remains of the same order as for MAPbI$_3$. The inhomogeneous contribution $W_G$ is reduced in MAPbBr$_3$, presumably due to the smaller size of the cage voids, which leads to a reduction of the reorientation possibilities for the MA molecules which, in turn, limits the spread of the bond-length distribution.

Interestingly, little temperature evolution of the Raman peaks is observed for MAPbCl$_3$ over the 90 to 300 K range, although similar broadening behaviour might be expected (see **Figure 8k to 8o and Figure S17c**). Only a few of the cage modes seem to be linearly broadened across the investigated temperature range and the Lorentzian nature of the Raman peaks is preserved across the phase changes. At low temperatures, the FWHM of the Raman peaks in MAPbCl$_3$ is large compared to the other compounds (at 90K the FWHM of a typical cage mode is ~ 10 cm$^{-1}$ for MAPbI$_3$, ~15 cm$^{-1}$ for MAPbBr$_3$ and ~40 cm$^{-1}$ in MAPbCl$_3$).

The same aforementioned studies[21, 22, 45] demonstrate that, for MAPbCl$_3$, the head-to-tail ordering of the cation in the orthorhombic phase is completely removed in the cubic phase. The limited temperature-evolution of the broadening in MAPbCl$_3$ can thus have two origins: (i) there is no reorientation of the MA cations in the cavity (coupled with a cage deformation mode), or (ii) these rearrangements do occur, but the initial and final surroundings of the cations are the same. NMR studies[44, 45] seem to rule out the first hypothesis. The chloride already adopts a cubic perovskite structure by 180 K. Of the three halides, the chloride structure tolerance factor is closest to the ideal value of 1.[53] We therefore propose that the unhindered rotations in the more cubic cavity might preserve the width of the Raman peaks, even at room temperature.

The conservation of Lorentzian shape of the Raman peaks on the investigated temperature range means that at all temperatures: $W_G \ll W_L$. This behaviour can again be rationalized by considerations on the size of the cage voids. On the one hand, the voids are sufficiently small that even though the reorientation of the MA molecules is unlocked in the cubic phase, inhomogeneous effects (i.e. $W_G$) are reduced because their emergence is limited by the highly symmetrical environment. On the other hand, the small size of the voids enhances a lot steric hindrance and consequently the dynamic coupling is enhanced as well. This leads to a strong increase in homogeneous linewidth which now dominates the Raman FWHM at any temperature.

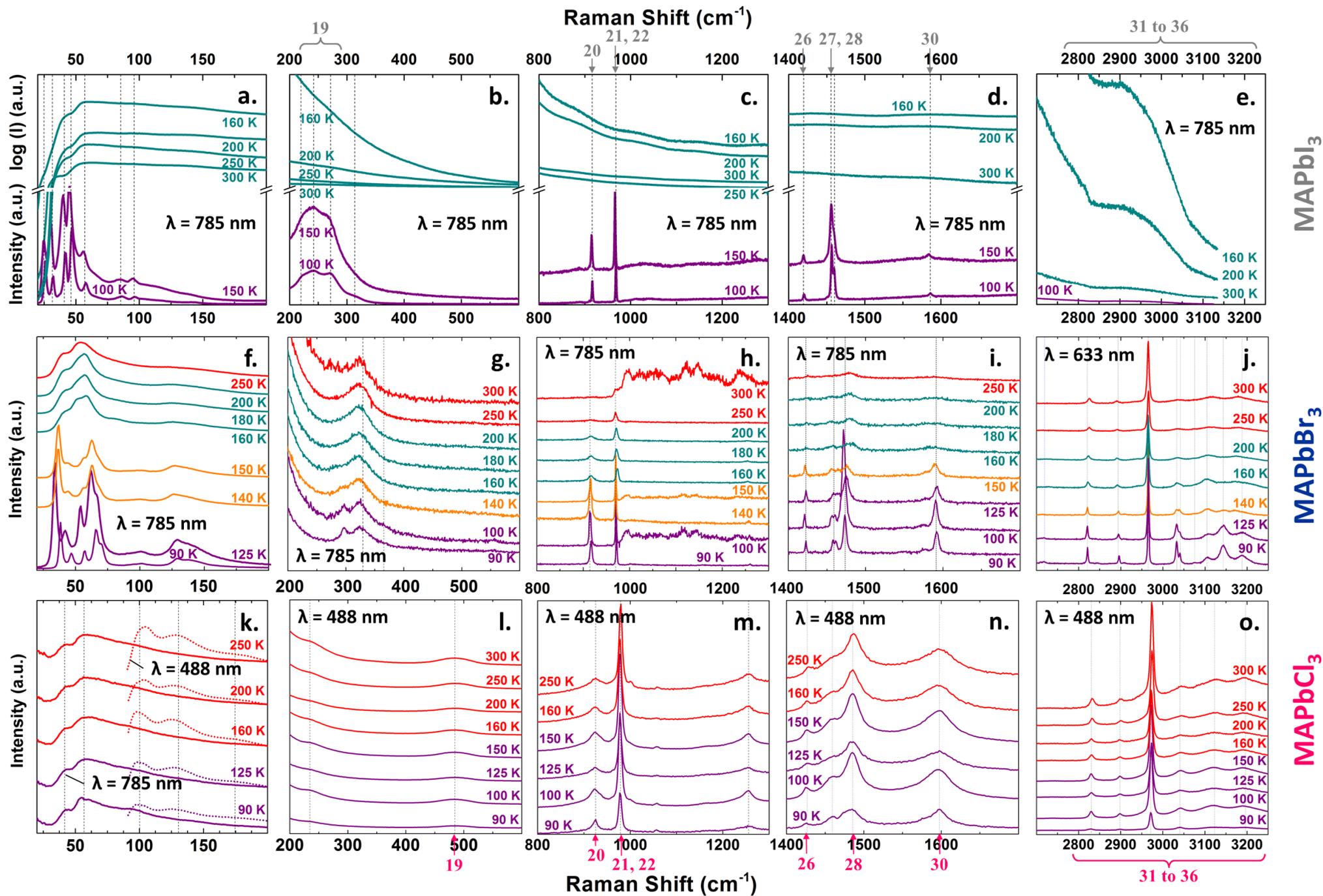

**Figure 8** Temperature dependence of the Raman spectra of MAPbI$_3$ (top row **a.** to **e.**), MAPbBr$_3$ (middle row **f.** to **j.**) and MAPbCl$_3$ (bottom row **k.** to **o.**). Each graph on a row has a different y-axis to best show the temperature evolution of the peaks. In the case of MAPbI$_3$, a logarithmic scale is used above the axis break to facilitate reading. The colour code is the same as in **Figure 1**. The data is reported for the excitation wavelength $\lambda$ yielding the clearest data (specified on each graph).

For all three compounds, the coupling between both organic and inorganic sub-lattices leads to an increase in the existing Raman peak widths rather than creating a distribution of new mode frequencies. The resulting homogeneous dynamic broadening of Raman peaks is inversely proportional to the associated phonon lifetime, $\tau$ ($\tau = \hbar$ / FWHM).[54] Thus the abrupt increase of most of the peak widths displayed in **Figure S17a** to **S17c** in the **supplementary information** corresponds to a drastic decrease of the phonon lifetime associated with the mode. The estimated typical phonon lifetimes of a cage mode (the 'nodding donkey' around N, mode 10 in the cubic phase), the C-N torsion mode (number 19 in the cubic phase), and a molecular vibration (the C-N symmetric stretch) are given for the three methylammonium lead halide compounds in **Figure 9a to 9c**. The data points plotted with open circles in **Figure 9** correspond to the estimated phonon lifetimes for homogeneously broadened peaks (*i.e.* peaks that are best fitted with Lorentzian functions). In that case, the Lorentzian contribution to the broadening can be deconvolved from the Gaussian contribution induced by the experimental setup, using the approach detailed in Ref. [55] and summarized in **supplementary note S3**. However, in the case when inhomogeneous broadening dominates, the former method can only yield a lower limit on the phonon lifetime (shown as filled square markers in **Figure 9**).

**Effect of vibrational modes on heat and electrical transport**

Introducing a weakly coupled mass inside a host lattice is a strategy to reduce its thermal conductivity. Efficient thermoelectric materials (i.e. with large values of *ZT*, the thermoelectric figure-of-merit) like skutterudites are designed with so called 'rattler' atoms.[56-58] Rattlers have been shown to be responsible for significant broadening of the Raman peaks compared with the "unfilled" structure, due to a drastic reduction of the phonon lifetimes.[59] Unlocking the reorientation of the MA ion in $CH_3NH_3PbX_3$ appears to act like the activation of a rattler. This is consistent with the recent observations showing that thermal conductivity is heavily supressed in $MAPbI_3$ compared with model systems,[60] and can decrease abruptly as the temperature is increased across the orthorhombic-to-tetragonal transition in films on some substrates.[61] The estimation of the thermal conductivity of $MAPbI_3$ in that study corroborates a previous experimental study[62] yielding a value between 0.3 and 0.5 $W.K^{-1}m^{-1}$ at room temperature. To put this value into perspective, it can be compared to the thermal conductivity of $PbI_2$ (~ 2.7 $W.K^{-1}m^{-1}$),[63] single layer graphene (~ 5000 $W.K^{-1}m^{-1}$),[64] Aluminium (~ 222 $W.K^{-1}m^{-1}$),[65] quartz glass (~ 1.4 $W.K^{-1}m^{-1}$),[66] the record thermoelectric materials CdTe and SnSe (~ 2 and 0.6 $W.K^{-1}m^{-1}$, respectively),[67, 68] the polymer P3HT (~ 0.2 $W.K^{-1}m^{-1}$)[69] or air (~ 0.025 $W.K^{-1}m^{-1}$).[70] It has been suggested that $CH_3NH_3AI_3$ (A = Pb and Sn) might be of interest for thermoelectric application, with $ZT$ values that could reach values between 1 and 2 (values above 2 are required to compete with conventional power generators).[71]

The phonon lifetimes seen in **Figure 9** (around 0.1 ps for $MAPbI_3$ or below) are generally short relative to other semiconductors at room temperature: for optical phonons at 300 K, values of ~2 ps and ~4 ps were reported in bulk silicon[72] and natural germanium crystals,[73] respectively. Lower values (between 0.1 and ~ 3 ps for optical phonons between 0.5 and 4 THz)[74] are expected at 300 K for the good thermoelectric materials PbSe and PbTe. In these materials, optical phonons contribute considerably to the lattice thermal conductivity and serve as important scattering channels for acoustic phonons.[74]

As discussed above, the short phonon lifetimes in $MAPbX_3$ imply a high degree of anharmonic phonon-phonon coupling. The phonon lifetimes observed, particularly at room temperature, where the Raman peaks can no longer be resolved for many modes, might also be expected to contribute to a reduction of the charge-carrier mobility through electron-phonon interactions. Two independent studies[75, 76] have reported that the temperature dependence of the mobility for $MAPbI_3$ is proportional to $T^{-1.4}$ or $T^{-1.6}$, which is close to the $T^{-1.5}$ behaviour expected for band-like transport limited by phonon scattering, as observed in germanium[77]. The mobility values reported for $CH_3NH_3PbX_3$ at room temperature – in excess of 100 $cm^2/Vs$ for $MAPbI_3$ (references [64, 65]) appear to be remarkably high for a solution processed material, particularly when combined with the observation that these materials display behaviour consistent with a high concentration of ionic defects.[3, 78, 79] However, mobilities up to 1000 $cm^2/Vs$ could be expected from the calculated electron band dispersion reported in several studies[80] that yield a charge carrier effective mass as small as $m_0^* \sim 0.12$ for electrons and 0.15 for holes.[81] At room temperature, two scattering processes may limit charge carrier mean free path in $CH_3NH_3PbX_3$: impurity scattering (independent of phonons) and electron-phonon scattering. Short phonon lifetimes indicate that inelastic electron-phonon scattering is likely, and energy transferred can dissipate quickly.

The optical phonon modes we have identified have much lower energies than conventional semiconductors, where the lowest optical modes are typically around 40 meV (9 THz).[54, 82] In these conventional materials the optical modes will not be significantly populated at room temperature, and since electrons will not have this energy either, these phonons will not contribute to electron scattering processes. This point is illustrated in **Figure 10**, this shows a factor proportional to electron-phonon (emission) scattering rate as a function of electron energy including a decomposition into the relative contributions to this rate from different modes. The figure indicates the sharp turn-on from the contribution of optical phonons to the electron-phonon scattering rate as electron energies increase. For $MAPbI_3$, we observe the lowest optical mode is activated from ~4 meV (1 THz), well below the thermal energy at room temperature (26 meV). This suggests that optical, rather than acoustic, phonon scattering may dominate loss of mobility at room temperature in these materials.

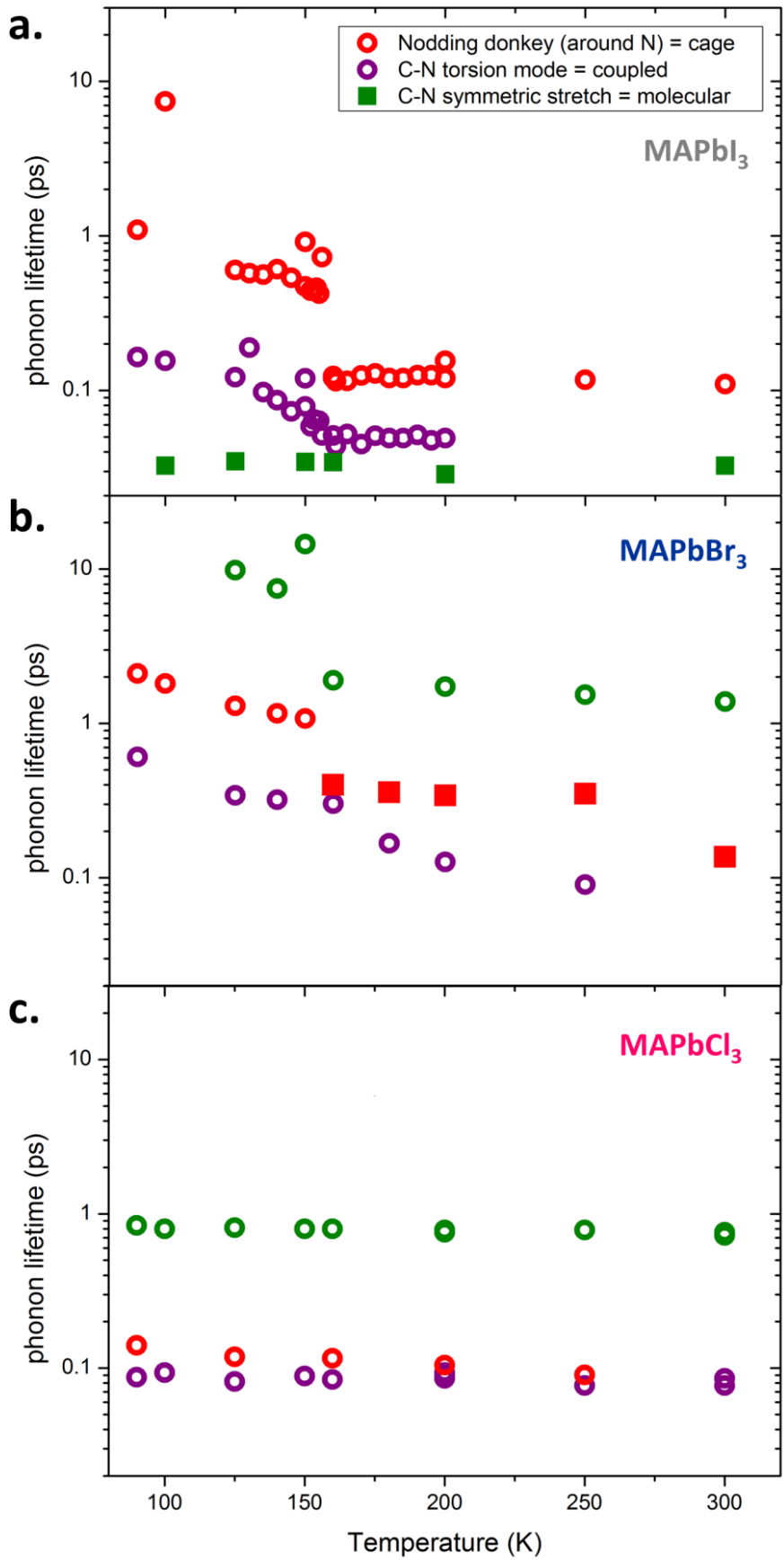

**Figure 9** Temperature dependence of the estimated phonon lifetimes for MAPbI$_3$ (**a.**), MAPbBr$_3$ (**b.**) and MAPbCl$_3$ (**c.**). Three representative phonons were selected: the green markers show a typical molecular mode for each compound (here the C-H symmetric stretch, see **Table 3**), the red markers stand for a typical cage mode (mode number 10, the 'nodding donkey' around N, see **Tables 1 and 2**), and the purple triangles represent the C-N torsion mode (i.e. the isolated mode, see **Table 3**). Open circle markers have been used to plot the lifetimes of homogeneously broadened peaks while filled square markers show the lower limit of the phonon lifetime in the case of inhomogeneous broadening, as detailed in **supplementary Note S3**.

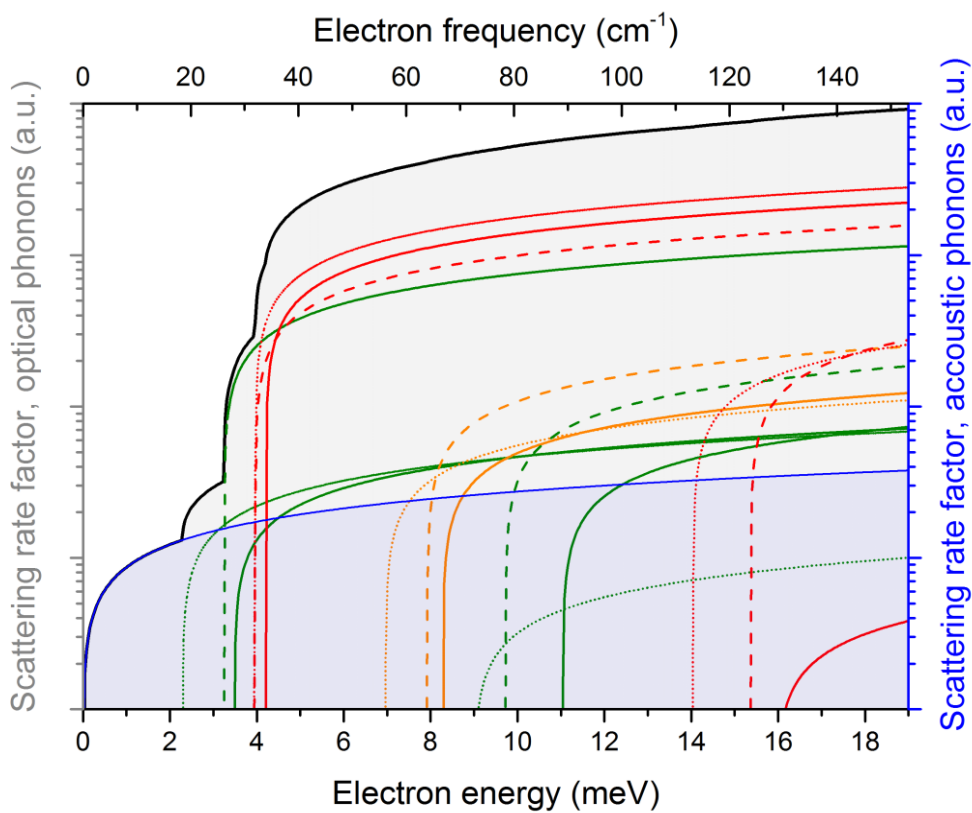

**Figure 10** Electron-phonon scattering rate factor for optical and acoustic phonons as a function of electron energy. This approximate quantity is synthesised from a superposition of estimated contributions from each phonon. These are derived from the energy of each mode at the Γ point, scaled by its relative IR activity (considered a proxy for its piezoelectric coupling), convolved with a parabolic density of electronic

states at finite temperature. The contribution of acoustic modes is obtained assuming equipartition holds.[83] Optical modes are assumed dispersion-less, acoustic modes with linear dispersion. Only optical phonon emission by electrons is accounted for in this figure, the contribution of phonon absorption processes by charge carriers can be found in reference [83] and would increase the scattering contribution of optical modes at lower energies. The sets of green, red and orange curves correspond to the optical eigenmodes in **Figure 3**. The black curve gives the sum of the optical phonon contribution to the scattering rate factor. The scattering rate per electron of a given energy at a temperature $T$ can be obtained by convolution of the scattering factor with a Bose-Einstein distribution function.

**Conclusion**

In conclusion we have performed a complete assignment of the main features observed in Raman and terahertz absorption spectra of the three methylammonium lead halide perovskites to their respective vibrational modes. From the temperature dependence of the spectra, we have shown the key role of two types of disorder in the $CH_3NH_3PbX_3$ material family, *viz.* (i) dynamic disorder caused by the unlocking of the rotation of the methylammonium ions in their cavities, causing homogeneous peak broadening, and (ii) statistical disorder caused by the various possible cation orientations, which leads to inhomogeneous peak broadening. In particular, we have demonstrated that statistical broadening can give rise to extra peaks in the spectra. The peak broadening occurring at the orthorhombic-to-tetragonal phase change in $MAPbI_3$ is the most pronounced and corresponds to a step-like decrease of the lifetimes of most of the low frequency modes. This change results in the observed decrease in the thermal conductivity of $MAPbI_3$ at room temperature. The observation of short phonon lifetimes combined with our assignment of low-energy optical phonons suggests that optical phonon scattering is likely to dominate at room temperature in these materials.

**Data Access**

The structures used for the phonon calculations are available from https://github.com/WMD-group/hybrid-perovskites, while the raw data from the phonon calculations, including simulated spectra and mode eigenvectors, are available from https://github.com/WMD-group/Phonons. Custom codes written to analyse gamma point motion are available at https://github.com/jarvist/Julia-Phonons, the customised version of Phonopy to provide eigenmode-resolved partial densities of states is available at https://github.com/jarvist/phonopy.

**Acknowledgments**

The authors thank Juraj Sibik and Axel Zeitler for their contribution in the measurement and interpretation of the terahertz absorption spectra. PB and AL are grateful to the EPSRC (EP/J002305/1, EP/M014797/1 and EP/M023532/1) for financial support. The work at Bath was supported by the EPSRC (EP/K016288/1, EP/K004956/1, EP/L000202, and EP/M009580/1) and the ERC (Grant 277757). ARG, MIA and MCQ thank the Spanish Ministry of Economy and Competitiveness (MINECO) for its support through Grant No. CSD2010-00044 (Consolider NANOTHERM) and MAT2015-70850-P (HIBRI2). The work at ICMAB was carried out under the auspices of the Spanish Severo Ochoa Centre of Excellence program (grant SEV-2015-0496). JN acknowledges the EPSRC for founding (EP/J017361). AP would like to thank the Royal Irish Academy for the Charlemont grant for funding the research visit that made the terahertz work possible.